\documentclass[12pt]{article}
\usepackage{epsfig} 

\newcommand{\etal}{\mbox{\it et al.}}
\newcommand{\nualpha}{\mbox{$\nu_{\alpha}$}}
\newcommand{\nui}{\mbox{$\nu_{\mathrm i}$}}
\newcommand{\ualphai}{\mbox{$U_{\mathrm \alpha i}$}}
\newcommand{\deltatwomnuij}{\mbox{$\Delta m^2_{\rm ij}$}}
\newcommand{\mtwonui}{\mbox{$m^2(\nu_{\mathrm i})$}}
\newcommand{\mtwonuj}{\mbox{$m^2(\nu_{\mathrm j})$}}
\newcommand{\mnui}{\mbox{$m(\nu_{\mathrm i})$}}
\newcommand{\mnuj}{\mbox{$m(\nu_{\mathrm j})$}}
\newcommand{\mnue}{\mbox{$m(\nu_{\mathrm e} )$}}
\newcommand{\mnumu}{\mbox{$m(\nu_{\mu} )$}}
\newcommand{\mnutau}{\mbox{$m(\nu_{\tau} )$}}
\newcommand{\mtwonue}{\mbox{$m^2(\nu_{\mathrm e} )$}}
\newcommand{\mtwoalpha}{\mbox{$m^2(\nu_{\alpha} )$}}
\newcommand{\mee}{\mbox{$m_{\mathrm ee}$}}

\newcommand{\mtwo}{\mbox{$m_\nu^2$}}

\newcommand{\ttwo}{\mbox{$\rm T_2$}}

\newcommand{\ev}{\mbox{$\rm eV/c^2$}}
\newcommand{\evtwo}{\mbox{$\rm eV^2/c^4$}}
\newcommand{\bdec}{\mbox{$\beta$~decay}}
\newcommand{\bspec}{\mbox{$\beta$~spectrum}}

\newcommand{\ezero}{\mbox{$E_0$}}

\title{THE NEUTRINO MASS DIRECT MEASUREMENTS}

\author{Ch. Weinheimer\\
  \normalsize Helmholtz-Institut f\"ur Strahlen- und Kernphysik,\\
  \normalsize Rheinische Friedrich-Wilhelms-Universit\"at,\\ 
  \normalsize D-53115 Bonn, Germany\\
  \normalsize Email: weinheimer@iskp.uni-bonn.de
}

\begin{document}

\maketitle

\begin{abstract}
One of the most important tasks in neutrino physics 
is to determine
the neutrino mass scale  to distinguish between hierarchical and 
degenerate neutrino mass models and to clarify the role of neutrinos 
as dark matter particles in the universe. 
The current tritium \bdec\ experiments at Mainz and Troitsk
are reaching their sensitivity limit. 
The different options for a next generation direct neutrino mass experiment
with sub-eV sensitivity are discussed. The KATRIN experiment,
which will investigate the tritium $\beta$ spectrum with an unprecedented 
precision, is being prepared to reach a  sensitivity of 0.2~eV.
\end{abstract}

\normalsize\baselineskip=15pt

\section{Introduction}
\label{sec_intro}  
Neutrinos are about 1 billion times more abundant in the universe than
baryons. Therefore already tiny neutrino masses of a few 
tenth~eV could contribute
significantly to the dark matter of the universe and influence structure
formation and the evolution of the universe.
Recent experimental results from atmospheric and solar neutrinos 
(see \cite{pdg02} and references therein) as well as from  
reactor neutrinos \cite{kamland03}
give strong evidence that neutrinos oscillate from one flavor
state into another.
Therefore, a neutrino of one
specific flavor eigenstate $\nualpha =  \sum_i \ualphai \nui$ is a non-trivial
superposition of neutrino mass states \nui , with at least two non-zero
neutrino mass values \mnui. 
Future oscillation experiments will determine the elements
\ualphai\ with great precision.

However, $\nu$--oscillations experiments do not yield the values
of the neutrino masses. In the case of pure neutrino 
vacuum oscillation they are only sensitive to differences
between squared neutrino masses $\deltatwomnuij =
|\mtwonui-\mtwonuj |$. The values $\deltatwomnuij$ from
oscillation experiments only give lower limits on neutrino masses
${max}\left(\mnui,\mnuj \right) \geq \sqrt{\deltatwomnuij}$.
On the other hand, if the absolute value of one  mass eigenstate
\nui\ is known, all other neutrino masses can be reconstructed
with the help of the differences of the squared neutrino masses (if 
the signs of the different $\mtwonui-\mtwonuj$ values are known).

Information on neutrino masses  can be inferred
by astrophysical observations and by laboratory experiments,
using two different approaches for the latter case: the
so-called ``direct mass measurements'' and the
search for  neutrinoless double \bdec . Both methods  give
complementary information on the neutrino masses \mnui\ as outlined 
in section 3.

Except time-of-flight measurements of neutrinos emitted in a 
  supernova the direct neutrino mass method is investigating
  the kinematics of weak decays. Here the charged decay products are
  measured and the
  missing neutrino mass is reconstructed from the kinematics of
  the charged particles by using energy and momentum conservation. 

  From its principle, 
  a kinematical neutrino mass measurement yields
  information on the different mass eigenstates \mnui , but usually the
  different neutrino mass eigenstates cannot be resolved by the
  experiment. Therefore for a measurement of a neutrino flavor $\nu_\alpha$ 
  an average over the neutrino mass eigenstates \nui\ contributing according
  to their mixing \ualphai\ is obtained:
  \begin{equation}
    \mtwoalpha =  \sum_{\mathrm i} |\ualphai^2| \cdot \mtwonui
  \end{equation}

The most sensitive information on a neutrino mass from direct 
mass experiments are the lowest upper limits of a few eV 
obtained for the mass of the electron 
neutrino by the investigation of the tritium \bdec . 
The present upper limits on the mass of the 
muon and tau neutrinos are $\mnumu < 190$~keV (90~\%~C.L.)
and $\mnutau < 18.2$~MeV (90~\%~C.L.) \cite{pdg02}.

This paper is organized as follows:
In section 2 the recent results of the tritium \bdec\ experiments
at Mainz and Troitsk are presented. In section 3 the motivation, 
the options and requirements for future neutrino mass 
measurements are discussed briefly. In section 4 the upcoming 
KArlsruhe TRItium Neutrino experiment KATRIN is presented. Section 5 gives
the conclusions.

\section{The Mainz and Troitsk tritium $\mathbf \beta$ decay experiments}
The Mainz and Troitsk tritium \bdec\ experiments are using both 
integrating $\beta$ electron spectrometers of MAC-E-Filter type, 
which provide  high luminosity and low background combined
with an energy resolution of 4.8~eV and 3.5~eV, respectively. 
Mainz uses a thin film of molecular tritium quench-condensed 
on a cold graphite substrate as tritium source, 
whereas Troitsk has chosen a windowless gaseous
molecular tritium source. After the upgrade of the Mainz experiment
in 1995-1997 both experiments are running with similar signal and
background rates.

\subsection{The Troitsk results}

\begin{figure}
\epsfxsize=14cm
\centerline{\epsffile{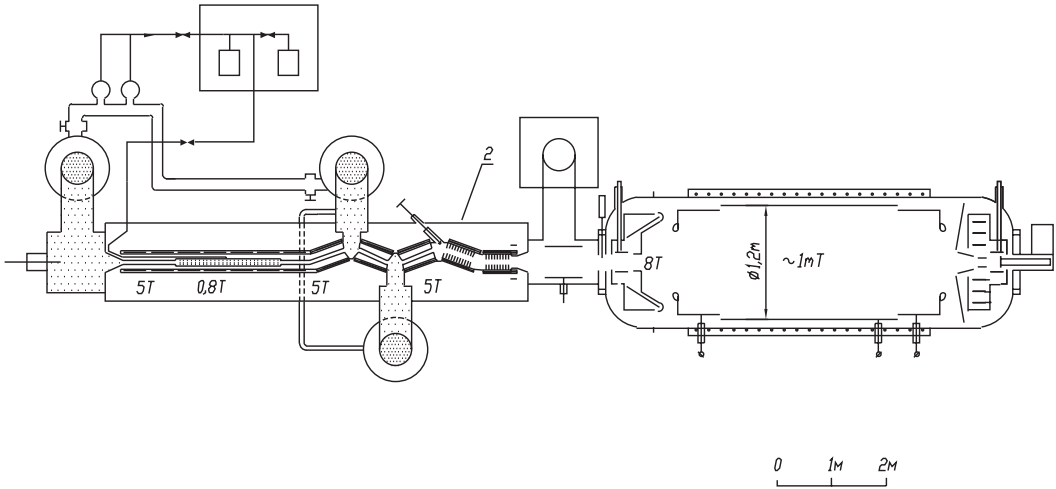}}
  \caption{Schematic view of the Troitsk experimental setup 
   \protect\cite{bel95}, 
   showing
   from left to right:the windowless gaseous tritium source of a 3~m long tube
    of 50~mm diameter, the differential and cryo-pumping
    section, 
    the integrating electrostatic spectrometer of MAC-E-Filter type.}
  \label{fig_troitsk_setup}
\end{figure}

Figure \ref{fig_troitsk_setup} shows the Troitsk neutrino mass experiment.
From its first data taking
in 1994 on the Troitsk experiment reports an anomalous excess in the
experimental \bspec\ appearing 
as a sharp step of the count
rate at a few eV below the endpoint of the \bspec\ \ezero\ \cite{bel95}.
Since the Troitsk spectrometer is integrating, this step 
corresponds to a line in the primary spectrum with a relative intensity of 
about $10^{-10}$ of the total decay rate.
Later the Troitsk group reported that the position of this line
oscillates with a frequency of 0.5 years between 5~eV and 15~eV below
\ezero\ \cite{lob99}. The Troitsk experiment is correcting for this 
anomaly by fitting an additional line to the \bspec\ run-by-run. 

Combining the 2001 results with the previous ones from 1994--1999
\cite{lob00} gives \cite{lob02}
\begin{equation}
\mtwonue = -2.3 \pm 2.5 \pm 2.0~ \evtwo 
\end{equation}
from which an upper limit on \mnue\ is obtained of 
\begin{equation}
\mnue < 2.05~ \ev \quad {\rm (95~\%~C.L.)}
\end{equation}
This limit is valid under the assumption that the anomalous excess count rate
near the endpoint is described by an additional line correctly.

\subsection{The recent Mainz data}

After the upgrade at Mainz runs of a total length of about 1 year have been taken
up to the end of 2001. From late 1998 on a high-frequency pulsing on one
of the electrodes was applied inbetween measurements every 20~s to lower
and stabilize the background. From that time on no indication of
any Troitsk-like anomaly was observed. Background instabilities 
did not allow to extract neutrino mass results for the 2000 data.
 
Additional studies on quench condensed 
\ttwo\ films clarified their energy
loss function \cite{eloss}, 
their self-charging \cite{bornb2002}, 
and their dewetting as a function of 
temperature \cite{fleischmann2}.

\begin{figure}[tb]
\centerline{\epsfxsize=6cm\epsfbox{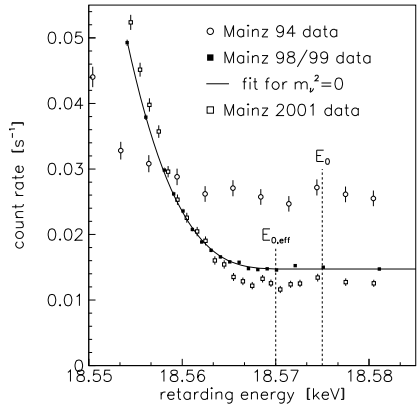}}
\caption{Averaged count rate of the 1998/1999 Mainz data
filled squares) with fit (line) 
and of the 2001 Mainz data (open squares) 
im comparison with previous Mainz data
from 1994 (open circles) \protect\cite{bonn97} 
as function of the retarding energy
$-eU$  near the endpoint \ezero , and effective endpoint $E_{0,eff}$. 
The position of 
the latter takes into account 
the width of response function of the setup and 
the mean rotation-vibration excitation energy of the electronic 
ground state of the $\rm ^3HeT^+$ daughter molecule.}
\label{fig_mainz_data}
\end{figure}

Fig. \ref{fig_mainz_data} shows the integral count rate averaged over the
1998/1999 and 2001 runs as 
function close to the endpoint \ezero = 18575~eV;
data obtained in 1994 \cite{bonn97} are shown for 
comparison.
The improved Mainz setup yields a signal-to-background ratio 10 times 
better than before and much better statistics has been obtained 
meanwhile.

\begin{figure}[hbt]
\centerline{\epsfxsize=6cm\epsfbox{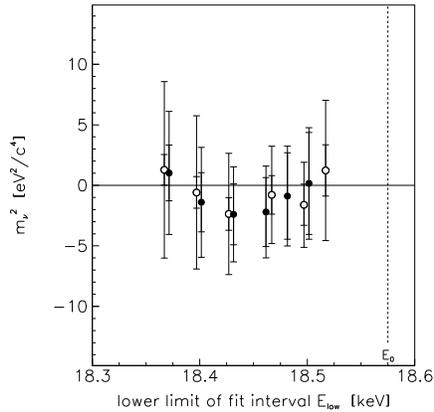}}
  \caption{Mainz fit results on \mtwonue\ as a function of the
  the lower boundary of the fit interval (the upper bound is fixed
  at 18.66~keV, well above \ezero ) for data from 1998 and 1999
  \protect\cite{bonn00} (open circles) and from the last runs of 2001
  (filled circles) \protect\cite{kraus02}.
  The error bars show the statistical uncertainties
  (inner bar) and the total uncertainty (outer bar).
  The correlation of data points for large fit intervals
  is due to the uncertainties of the systematic corrections, which are
  dominant for fit intervals with a lower boundary E$_{low}$\,$<$~18.5~keV.}
  \label{fig_mainz_fits}
\end{figure}

 Figure \ref{fig_mainz_fits} shows
 the fit results on \mtwo\ with statistical  
 and total uncertainties 
 for the 4 different runs Q5 to Q8 of 1998/1999 and of Q11 and Q12 of 2001
 as function
 of the lower energy limit of the data interval used for the analysis.
  The monotonous trend towards negative values of \mtwo\ for larger
  fit intervals as it was observed for the Mainz 1991 and 1994 data 
  \cite{bonn97} has vanished. This shows
  that the dewetting of the \ttwo\ film from the graphite substrate 
  \cite{fleischmann2} indeed was the reason for 
  this behavior. Now this effect is safely suppressed at the much
  lower temperature of the \ttwo\ film.   
Moreover, the neutrino squared masses obtained from the fit are very stable
and compatible with zero within their uncertainties and the
previous Mainz results (see figure \ref{fig_mainz_fits}). No indication
of a Troitsk-like anomaly or any residual problem in the Mainz
data were found.

For the data set of 1998 and 1999 the energy interval of the 
last 70~eV below the endpoint 
the combined statistical and systematic uncertainty 
attains a minimum. The result for \mtwo\  is \cite{bonn00}
\begin{equation}
  \label{eq_m2_9899}
  \mtwo   =  -1.6 \pm 2.5 \pm 2.1~\evtwo\,. 
\end{equation}

 The result for the 2001 data of the last 70~eV of
the \bspec\ below the endpoint (E$_{\rm low}$=18.5~keV, see fig. 
\ref{fig_mainz_fits}) on \mtwonue\ is \cite{kraus02}:
\begin{equation}
\mtwonue = +0.1 \pm 4.2 \pm 2.0~ \evtwo 
\end{equation}
Combining this value with the one obtained from the data sets
from 1998 and 1999 (\ref{eq_m2_9899}) gives
\begin{equation}
\mtwonue = -1.2 \pm 2.2 \pm 2.1~ \evtwo 
\end{equation}
which corresponds to an upper limit \cite{kraus02} of
\begin{equation}
\mnue < 2.2~ \ev \quad {\rm (95~\%~C.L.)}
\end{equation}

The inclusion of the high-quality data from 2001 improves the
Mainz sensitivity only marginally, showing that the Mainz
experiment has reached its sensitivity limit.

In spring 2002 the Mainz group has installed a new
electrode system to check new ideas to avoid background and to
remove trapped particles. First measurements, performed after
this conference, showed that the new
ideas indeed are reducing the background rate by nearly a factor of 3.

\section{Future direct neutrino mass searches}
\label{sec_future}
The compelling evidence for non-zero
neutrino masses from atmospheric, solar and reactor 
neutrino experiments -- as briefly discussed in section 1 -- provides 
squared neutrino mass differences but no absolute neutrino masses.
This fact clearly demand
for the determination of the absolute neutrino mass scale as one of the
most important next steps in neutrino physics since the absolute neutrino
mass has strong consequences for astrophysics and cosmology as well as for
nuclear and particle physics: The neutrino mass states can be arranged in a
hierarchical way like the charged fermions. 
This would mean that
the different neutrino masses are
essentially governed by the square roots of $\deltatwomnuij$. 
On the other hand the neutrino masses could be quasi-degenerate 
with about the same value -- {\it e.g.} a few tenth of an eV -- and
small mass differences between the different states to explain 
the oscillation
signal. The latter case would be very important for cosmology (concerning
structure formation, evolution of the universe, \dots ) the former 
one much less. Both
scenarios
would require different expansions of the Standard Model of
particle physics to include these neutrino masses. 

The various ideas and approaches 
to determine the absolute neutrino mass with
sub-eV sensitivity  will be briefly discussed in the following:
\vspace*{-0.2cm}

\subsection{Cosmic microwave background radiation and large scale structure}
The observation of the structure in the universe at different scales and
the angular distribution of the fluctuations of the cosmic microwave
background radiation allows to set constraints on the Hot Dark
Matter content of the early universe and because of the 
relic neutrino density of about 112 neutrinos per flavour and cm$^3$
on the neutrino mass. 
Although very recently the WMAP experiment \cite{WMAP} reports 
on an upper limit of the masses of all neutrino of 0.7~eV,
the results derived this way is model dependent (a more conservative
approach gives limits of 1~eV or 2~eV, respectively \cite{hannestad}).
The model dependence is clearly illustrated by the fact, that Allen \etal\
obtain a non-zero sum of all neutrino masses of 0.64~eV from nearly
the same data by changing some assumptions on mass fluctuations 
(different amplitude of mass fluctuations on $8 h^{-1}$~Mpc scales
$\sigma_8$) \cite{allen}.
Additionally, there are strong degeneracies between the different 
astrophysical 
parameters and it is therefore very helpful to bring in information from
laboratory neutrino mass experiments to determine the other astrophysics
parameters more precisely. Last but not least, one should not forget, that
the present cosmological model depends on yet non-understood ``Dark Energy'', 
non-identified Cold Dark Matter and non-understood inflation, therefore,
the laboratory measurement of the neutrino mass scale could serve
as an important check of standard cosmology.
\vspace*{-0.2cm}

\subsection{Time-of-flight of supernova neutrinos}
Due to the smallness of neutrino masses 
the only laboratory to measure them by time-of-flight 
is our universe.
The correlation between energy and arrival time on earth 
of supernova neutrinos depends on their mass, 
thus allowing to extract the neutrino mass
by measuring arrival time and energy. 
Although a  supernova, exploding within our galaxy, would give hundreds to 
thousands of neutrino events in the current underground neutrino 
detectors,
the systematic uncertainty
connected with the not precisely known neutrino emission time spectrum does
not allow a sub-eV sensitivity on the neutrino mass.
\vspace*{-0.2cm}

\subsection{ Neutrinoless double \bdec }
The neutrinoless double \bdec\ is sensitive to the so-called
  ``effective'' neutrino mass
  \begin{equation}
    \label{eq_mee}
    \mee = | \sum_{\mathrm i} U_{\mathrm ei}^2 \cdot \mnui |\quad ,
  \end{equation}
  which is a coherent sum over all mass eigenstates contributing to the 
  electron neutrino.
  The determination of \mee\ from the measurement of the neutrinoless
  double \bdec\ rate is complementary to the direct determination
  of the mass of the electron neutrino since \mee\ and \mnue\ can
  differ by the following reasons:
  \begin{enumerate}
    \item Double \bdec\ requires the neutrino to be a Majorana particle.
    \item In the notation of eq. (\ref{eq_mee}) the 
       values $U_{\mathrm ei}^2$ can have -- in addition to a possible
       complex phase from  the $ 3\times 3$ neutrino mixing --
       two non-trivial so-called Majorana phases. This 
        could lead to a partial cancellation of the
       different terms of the sum. Especially that the recent solar neutrino
       data point to large mixing opens this possibility
       \cite{smirnov}.
    \item The uncertainty of the nuclear matrix elements
       of neutrinoless double \bdec\ still contributes to the 
       uncertainty of \mee\ by about a factor of 2 .
     \item Non Standard Model processes, others than the exchange
       of a Majorana neutrino, could enhance the observed neutrinoless
       double \bdec\ rate without changing \mee . 
  \end{enumerate}
The proposed double \bdec\ experiments of the next generation 
aim for a sensitivity
on \mee\ in the range of 0.1~eV and below \cite{elliott_vogel}.
\vspace*{-0.2cm}

\subsection{ Rhenium cryogenic bolometer experiments}
A straightforward approach to directly measure the electron neutrino
mass is the use of cryogenic bolometers.  This new
technique has been applied to the isotope $^{187}{\rm Re}$, which
has with $\ezero = 2.5$~keV 
the lowest $\beta$ endpoint energy and which optimizes the interesting
fraction below the endpoint \cite{Milano,gatti}. The
experiments are still in the early stage of development. Current
Rhenium micro-calorimeters reach an energy resolution of $\Delta E \sim
30$\,eV  \cite{Milano} and yield an upper limits of 22\,eV\ and 26\,eV,
respectively \cite{Milano,gatti}. To further improve the statistical
accuracy the operation of large arrays of
micro-calorimeters with better resolution is required. New techniques
are explored to enable these improvements.
The expected sensitivity on \mnue\
in the future is in the eV region \cite{gatti}.
\vspace*{-0.2cm}

\subsection{ The KATRIN experiment}
Summarizing the discussion above clearly means that one or more 
next generation double \bdec\ experiments have to be performed due to
their very low sensitivity. But considering the complementariness 
of neutrinoless double \bdec\ and the direct neutrino mass determination
it is also
clear that a next generation direct mass search has to be done. None of the
alternative direct methods discussed above is able to provide 
a sub-eV sensitivity in a model independent way within the next decade. 
Therefore, it is straightforward
to explore which sensitivity could be achieved by
investigating the tritium \bdec\ spectrum near its endpoint 
with the very successful 
MAC-E-Filter as spectrometer.

Discussions between groups from Mainz, Karlsruhe and Troitsk led to the 
proposal for a next generation tritium \bdec\ experiment to be built at
Forschungszentrum Karlsruhe/Germany. 
Now a strong collaboration including nearly the complete
worldwide expertise on tritium \bdec\ neutrino mass experiments has come
together and has published a 
Letter of Intent for the KATRIN experiment (KArlsruhe TRItium Neutrino
experiment) \cite{katrin_loi}.

\section{The KATRIN experiment}

\begin{figure}[tb]
\centerline{\epsfxsize=14cm\epsfbox{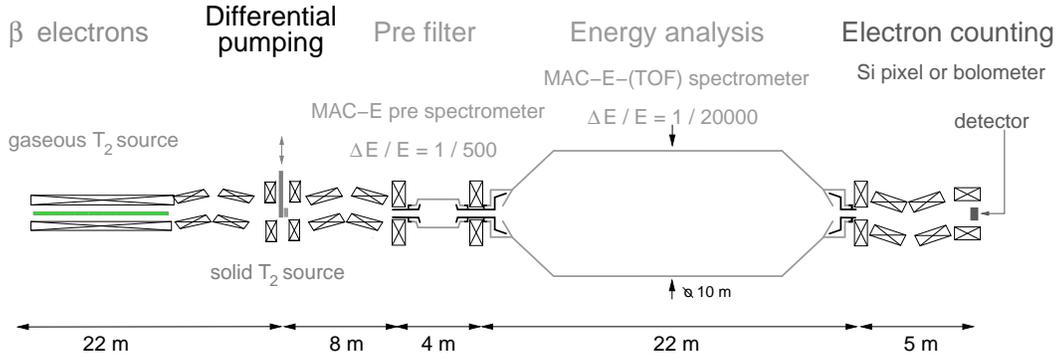}}
\caption{Schematic view of the proposed next-generation tritium
  \bdec\ experiment KATRIN. The main components of the
  system comprise a windowless gaseous tritium source (WGTS), a
  alternative quench condensed tritium source (QCTS), a pre-spectrometer,
  a large electrostatic
  spectrometer with an energy resolution
  of 1\,eV and a detector. 
   An electron transport system guides electrons from the
  \ttwo\ sources to the spectrometers, while eliminating all
  tritium molecules.
  \label{fig_katrin}}
\end{figure}

The KATRIN collaboration has enlarged and improved the proposed setup compared
to its Letter of Intent \cite{katrin_loi}. Figure 
 \ref{fig_katrin}) shows a schematic view of the proposed experimental 
 configuration.
The windowless gaseous tritium source (WGTS) allows the measurement of the
 endpoint region of the tritium \bdec\ and consequently  the determination
 of the neutrino mass with a minimum of systematic uncertainties 
from the tritium
 source. The WGTS will consist of a 10\,m long cylindrical tube of 90\,mm
diameter filled from the middle with \ttwo\ gas, resulting in a
source column density of about $(\rho d) \approx 5\cdot 10^{17}$ 
molecules/cm$^2$. 
With these values the count rate is increased by two orders of magnitude
with respect to the Troitsk experiment.
A  quench condensed tritium source (QCTS)
following the source concept of the Mainz experiment is considered as
a second alternative source, which has complementary systematics.

The electron transport system adiabatically guides \bdec\
electrons from the tritium sources to the spectrometer 
while at the same time eliminating any tritium flow towards
the spectrometer, which has to be
kept practically free of tritium for background reasons.
The first part of the transport system consists of a differential
pumping section with a tritium reduction of a factor $10^9$, the second part
of a liquid helium cold cryo-trapping section.

Between the tritium sources and the main spectrometer
a pre-spectrometer of MAC-E-Filter type will be inserted, acting
as an energy pre-filter to reject all $\beta$ electrons except the
ones in the region of interest close to the endpoint \ezero .
This minimizes the chances of causing background by ionization of
residual gas in the main spectrometer.
As the designs of the
pre- and main spectrometer will be similar, the former
is acting as a test facility for the larger main spectrometer. 
The design
and construction of the pre-spectrometer has already started.

 A key component of the new experiment will be the large
 electrostatic main spectrometer with a diameter of 10\,m and an overall
 length of about 22\,m. This high resolution MAC-E-Filter will allow
 to scan the tritium \bdec\ endpoint with increased
 luminosity at a resolution of a little bit less than 1\,eV, 
which is a factor of 4
 better than present MAC-E-Filters at Mainz and Troitsk. The 200 times
larger analyzing plane  with respect to the Mainz experiment allows
the remaining factor 50 to be utilized to increase the source cross section 
and, correspondingly, the signal rate.

The detector requires high efficiency
for electrons at $\ezero = 18.6$~keV and low $\gamma$ background. A high energy
resolution of  $\Delta E<600$\,eV for 18.6\,keV electrons should suppress
background events at different energies.
The present concept of the detector is based on a large array of
about 1000 silicon drift detectors 
surrounded by low-level passive shielding and
an active veto counter to reduce background.

At the International Workshop on Neutrino Telescopes already the proposed
enlarged version of the KATRIN experiment comprising a WGTS with 90~mm 
diameter and a main spectrometer with 10~m diameter was presented and 
consequently a sensitivity of 0.25~eV was reported.
The very recent simulations for 3 years of data taking using a
new strategy of optimized measurement point distribution 
and improved systematics
result in a neutrino mass sensitivity of even below 0.2~eV, 
with statistical and
systematic uncertainties contributing about equally. This sensitivity
number corresponds to an upper limit on the neutrino mass with 90~\% C.L., if 
no neutrino mass would be seen. To the contrary, 
a non-zero neutrino mass of 0.35~eV would be
detected with 5~$\sigma$ significance. 
This sensitivity improves
the existing limits by one order of magnitude and also
demonstrates the discovery potential of KATRIN for an electron
neutrino mass of a few tenth of an eV.

\section{Summary}
The current tritium \bdec\ experiments at Mainz and Troitsk
are reaching their sensitivity limits. The recent Mainz data have 
strictly the shape of a \bspec\ with zero neutrino mass, resulting in an upper 
limit on \mnue\ of 2.2~eV
at 95~\%~C.L.

A laboratory neutrino 
mass determination with sub-eV sensitivity is clearly needed
to distinguish between hierarchical and degenerate neutrino mass models and
to check the role of neutrinos in the early universe.
The search for the neutrinoless double \bdec\ is one very important approach.
Complementary and equally important is a next
generation direct neutrino mass experiment. Discussing the different options
shows that this experiment has to be a large tritium \bdec\ experiment
using a MAC-E-Filter. Such an experiment is being prepared 
by the KATRIN collaboration aiming for an sensitivity on the neutrino mass
of below 0.2~eV.

\section{Acknowledgments}
The author would like to thank the various 
collaborations for providing him kindly the presented informations and results.
The work of the Mainz and the KATRIN experiments connected to the
author is supported by 
the German Bundesministerium f\"ur Bildung und Forschung under
contracts 06MZ866I/5 and 05CK2PD1/5.

\end{document}